\documentclass[prb,aps,twocolumn,amsmath,amssymb,superscriptaddress,citeautoscript]{revtex4-1}
\usepackage[utf8]{inputenc}
\usepackage[english]{babel}
\usepackage{graphicx}  
\usepackage{dcolumn}   
\usepackage{bm}        
\usepackage{verbatim}   
\usepackage{color}		

\makeatletter
\newcommand*{\citenums}[2][]{%
	\begingroup
	\let\NAT@mbox=\mbox
	\let\@cite\NAT@citenum
	\let\NAT@super@kern\relax
	\renewcommand\NAT@open{}%
	\renewcommand\NAT@close{}%
	\cite[#1]{#2}%
	\endgroup
}
\makeatother

\begin{document}

\title{Dynamics of the Optical Spin Hall Effect}
\author{Daniel Schmidt}
\author{Bernd Berger}
\affiliation{
	Experimentelle Physik 2,
	Technische Universit\"at Dortmund,
	D-44221 Dortmund, Germany
}  
\author{Manfred Bayer}
\affiliation{
	Experimentelle Physik 2,
	Technische Universit\"at Dortmund,
	D-44221 Dortmund, Germany
}
\affiliation{
	A. F. Ioffe Physical-Technical Institute,
	Russian Academy of Sciences,
	St Petersburg 194021, Russia
}
\author{Christian Schneider}
\author{Martin Kamp}
\affiliation{
	Technische Physik,
	Universit\"at W\"urzburg,
	97074 W\"urzburg, Germany
}
\author{Sven H\"ofling}
\affiliation{
	Technische Physik,
	Universit\"at W\"urzburg,
	97074 W\"urzburg, Germany
}
\affiliation{
	SUPA, School of Physics and Astronomy, University of St Andrews, St Andrews, KY16 9SS, United Kingdom
}

\author{Evgeny Sedov}
\affiliation{
	Department of Physics and Applied Mathematics,
	Vladimir State University named after A. G. and N. G. Stoletovs,
	Gorky str. 87, 600000, Vladimir, Russia
}
\affiliation{
	School of Physics and Astronomy,
	University of Southampton,
	SO17 1NJ Southampton, United Kingdom
}

\author{Alexey Kavokin}
\affiliation{
	School of Physics and Astronomy,
	University of Southampton,
	SO17 1NJ Southampton, United Kingdom
}
\affiliation{
	Spin Optics Laboratory,
	St. Petersburg State University,
	Ul’anovskaya 1, Peterhof, St. Petersburg 198504, Russia
}

\affiliation{
CNR-SPIN, Viale del Politecnico 1, I-00133, Rome, Italy
}

\author{Marc A\ss mann}
\affiliation{
	Experimentelle Physik 2,
	Technische Universit\"at Dortmund,
	D-44221 Dortmund, Germany
}  
\date{\today}
\begin{abstract}
We study the time evolution of the Optical Spin Hall Effect (OSHE), which occurs when exciton-polaritons undergo resonant Rayleigh scattering. The resulting spin pattern in momentum space is quantified by calculating the degree of circular polarization of the momentum space image for each point in time. We find the degree of circular polarization performing oscillations, which can be described within the framework of the pseudospin model by Kavokin et al. (Ref. \citenums{kavokin2005}).
 \end{abstract}

\maketitle

\section{Introduction}
Exciton-polaritons, which are composite bosons arising from strong coupling between quantum well (QW) excitons and cavity photons, play an important role in the rapidly developing research field of spin-optronics or semiconductor spin optics.
Due to their excitonic component polaritons interact with each other, whereas the photonic component allows much faster propagation compared to bare excitons. The resulting optical reconfigurability of exciton-polaritons in semiconductor microcavities therefore is a striking feature which potentially allows one to realize all-optical devices for information technology. A lot of phenomena related to polariton switching and propagation recently have been investigated experimentally, such as polarization dependent switching\cite{amo2010}, all-optical logic gates\cite{leyder2007,ballarini2013}, bistable\cite{sarkar2010} and multistable \cite{vishnevsky2012,quellet_plamondon2016} switching of polariton reservoirs.
Also the propagation dynamics of exciton-polaritons has been studied extensively\cite{freixanet2000propagation,adrados2011,wertz2012,langbein2007}.
\\ \indent One possible way of creating polariton flows is by use of the non-resonant optical excitation energetically far above the resonant cavity mode. Using tailored spatial excitation patterns all-optical flow control\cite{schmutzler2015}, polariton amplification \cite{niemietz2016} and advanced momentum control\cite{assmann2012} have been demonstrated successfully. Polariton flows can also be created by resonant optical excitation of the lower polariton branch. The momentum direction of the in-plane propagation is then directly controlled by the incidence angle of the excitation beam. When a polariton flow is directed onto a defect in the sample, such as a structural defect or impurity, a scattering ring builds up in momentum space due to elastic scattering\cite{houdre2000,langbein2002,freixanet2000}. 
\\  \indent The polaritons inherit fundamental properties from their constituents.
Heavy-hole excitons are characterized by angular momentum projections on the QW growth axis that take values $\pm1$ and $\pm2$.
The so called dark excitons with $\pm2$ projections are optically inactive and in most cases their influence on light-matter interaction processes can be safely neglected.
At the same time, the bright excitons with angular momentum projections $\pm1$ are allowed to couple with cavity photons of two opposite (right and left) circular polarizations, respectively, and form a polariton spin doublet state. 
To describe this doublet, we can introduce the pseudospin vector $\mathbf{S}$ whose behavior in time is governed by the built-in effective magnetic fields of different origin.
\\ \indent In this work, we investigate the dynamics of the optical spin Hall effect which takes place due to the elastic scattering processes in a spin-polarized gas of propagating exciton polaritons\cite{kavokin2005}.
The OSHE describes the formation of spin currents due to an effective magnetic field, which is induced by the longitudinal-transverse (LT) splitting of the cavity photon modes.
In the passage of time characteristic anisotropic spin patterns form in the momentum space.
The recent works reported on the observation of the spin currents\cite{leyder2007observation,maragkou2011} and anisotropy effects\cite{Amo2009c} in both real and momentum space. Further, a non-linear analogue of the OSHE was demonstrated\cite{kammann2012}. Even the influence of an external magnetic field was examined theoretically\cite{morina2013}. All these works demonstrate the control of polariton spin currents, especially in the time-integrated momentum space.\\
\indent In the present work we aim for gaining control over the spin currents in the time domain. Therefore, we perform time resolved measurements of the momentum space of elastically scattered polaritons. We identify the spin currents induced by the OSHE and experimentally study their time dependence. The measured temporal behavior can be well reproduced utilizing the pseudospin model\cite{kavokin2004,kavokin2005}.

\section{Experiment}
\begin{figure}
\includegraphics[width=0.5\textwidth]{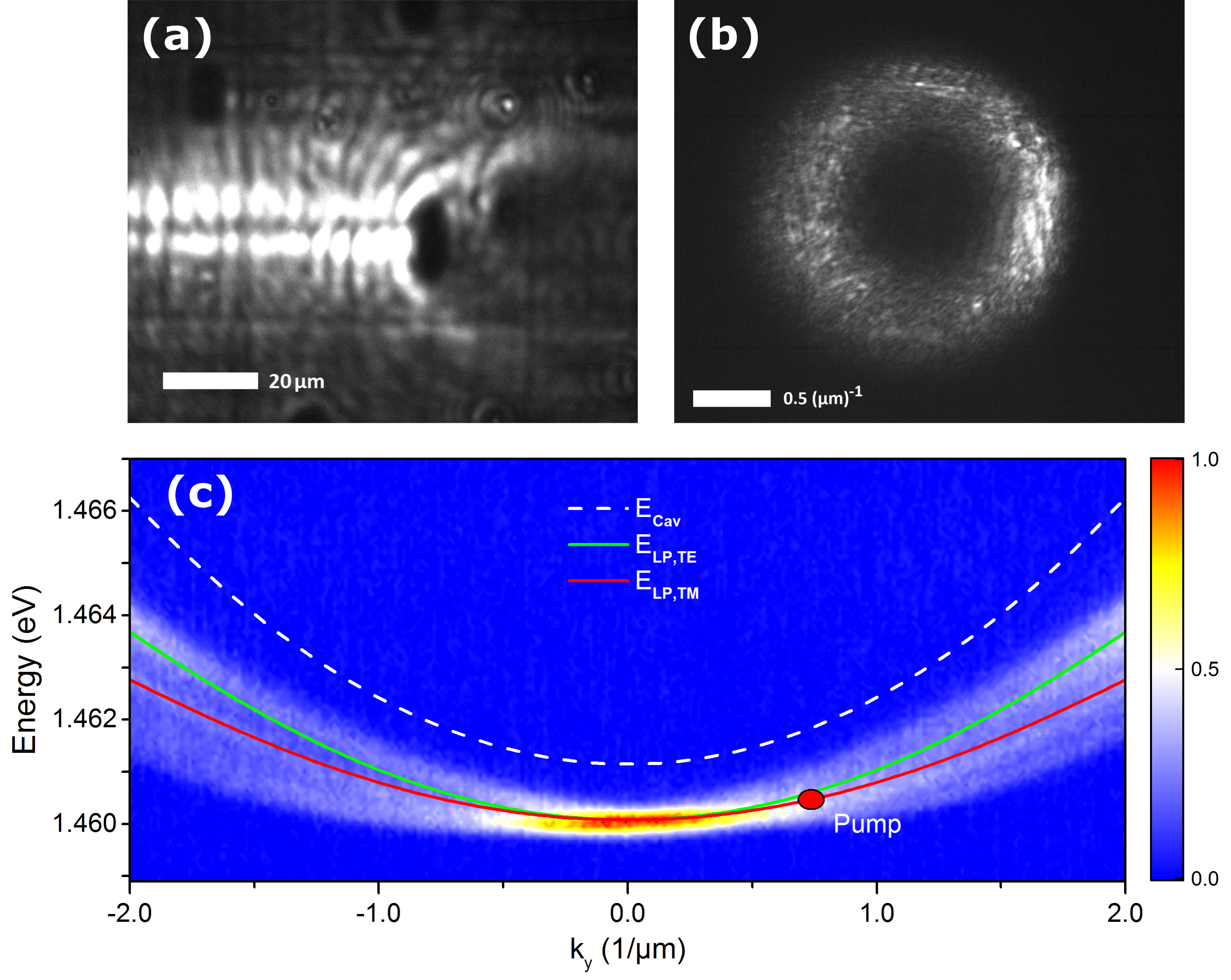}
\caption{(a) Real space image of resonant scattering (b) Corresponding momentum space image of resonant scattering. (c) Dispersion probed by non-resonant excitation with $k=0$. The TE-TM splitting is clearly visible. The pump configuration used when performing resonant scattering experiments is marked with the red dot. The polarization is matched to the TM-mode of the lower polariton branch. The entrance slit of the spectrometer is aligned central to the emission spot.}
	\label{fig:depicted_excitation}
\end{figure}

Our sample is a planar GaAs $\lambda$-cavity with six $\text{In}_{0.1}\text{Ga}_{0.9}$As quantum wells placed at the central antinodes of the confined light field and 26 top and 30 bottom GaAs/AlAs DBR layer pairs. The Rabi splitting amounts to about 6\,meV.
The measurements are performed at the detuning $\delta=E_\text{cav}(k_{||}\text{=0})-E_\text{exc}(k_{||}\text{=0})$ of $-5.3$~meV. We excite the sample resonant to the TM-mode of the lower polariton branch with a pulsed picosecond Ti:Sa laser centered at $848.86$~nm (1.46060~eV) and create polaritons with an in-plane wave vector of $k_{||}=0.73\,(\mu \text{m})^{-1}$. The power is 15~mW with a Gaussian spot diameter of $20$~$\mu \text{m}$. Using a cold-finger continuous flow cryostat the measurements are performed at 15~K. For spectral analysis the signal is sent to a $500$~mm monochromator equipped with a nitrogen cooled CCD camera. A Hamamatsu streak camera with S-20 photocathode allows for time-resolved measurements, for which an interference filter ensures that only the emission from the lower polariton branch (LPB) is detected.
\\  \indent See Fig. \ref{fig:depicted_excitation} for the intermediate images of the real space (a) and momentum space (b). For comparison the whole dispersion of the polaritons following non-resonant excitation is shown in Fig. \ref{fig:depicted_excitation} (c).
In the momentum space for certain sample positions a typical scattering ring appears, due to the elastic Rayleigh scattering of exciton polaritons by sample inhomogeneities or defects. Under resonant excitation and non-zero incidence pronounced defect scattering can be observed, see Fig. \ref{fig:depicted_excitation} (a). The propagating exciton-polaritons undergo self interference resulting in a typical standing wave pattern around the defect.
\\  \indent Here we temporally resolve this scattering process. Further, the distribution of spins is investigated to evaluate the pseudospin precession due to the OSHE. 
Therefore we polarize the incident beam linearly and direct it onto the sample under an aligned angle to resonantly excite the TM-mode of the LPB.
\\  \indent Since we want to capture the full time evolution of the momentum space distribution, we record single frames with $k_x$-$t$ information and merge them. To do so, we move the endmost lens in front of the streak camera stepwise to shift the momentum space image along the $k_y$-direction, perpendicular to the entrance slit of the streak camera \cite{nardin2009}. Thereby we incrementally capture the full time evolution of the sca\-t\-te\-ring ring in momentum space for both $\sigma^+$ and $\sigma^-$ polarization. See the supplementary information for a video of the pseudospin evolution in momentum space. 

\section{Results}
\begin{figure}
	\includegraphics[width=0.45\textwidth]{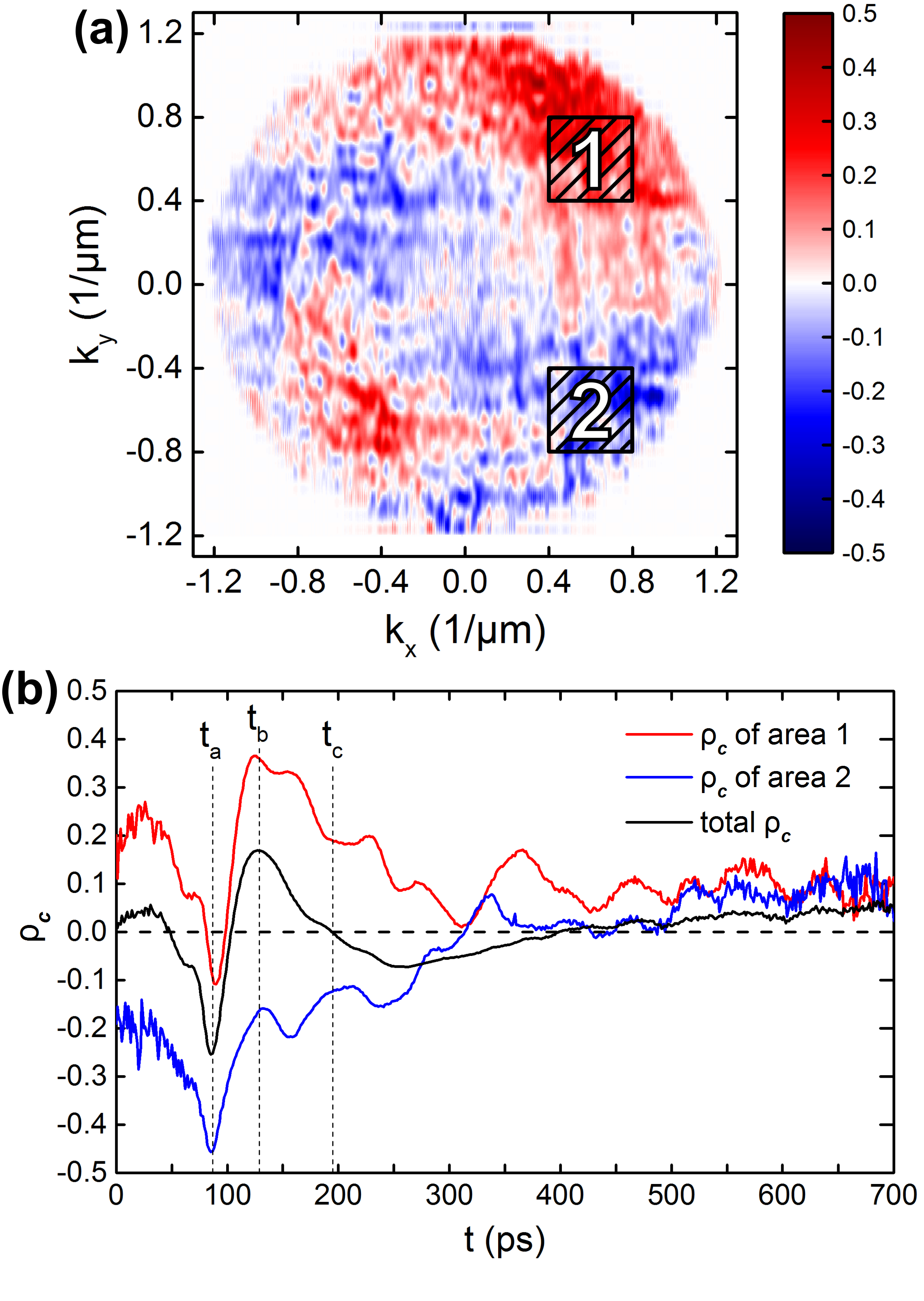}
	\caption{(a) Time integrated image of $\rho_c$ in momentum space with marked areas 1 and 2 (b) Time evolution of $\rho_c$ in selected areas 1 and 2. The recorded time frame starts at $t_0=0$ prior to the incidence of the excitation pulse.}
	\label{fig:dcp_time_integration_and_evolution}
\end{figure}

From the recorded data, the degree of circular polarization $\rho_\text{c}$ can be extracted at each $k_x$-$k_y$-point in momentum space for any point in time using
\begin{equation}
	\rho_\text{c} = \frac{I_{\sigma^+} - I_{\sigma^-}}{I_{\sigma^+} + I_{\sigma^-}}.
	\label{eq:DCP}
\end{equation}
We integrate the data over time and calculate $\rho_c$ to verify the typical pseudospin pattern of the OSHE. This is shown in Fig. \ref{fig:dcp_time_integration_and_evolution}(a). To reveal the time evolution of $\rho_c$ two areas are selected at angles of $+45^\circ$ and $-45^\circ$ with respect to the direction of the pump. Results for the chosen areas are shown in Fig. \ref{fig:dcp_time_integration_and_evolution}(b), where oscillations with opposite signs of $\rho_c$ can be observed. This resembles the behavior [\citenums{kavokin2005}] originally predicted. However, some striking deviations can be observed. First, $\rho_c$ in both areas initially drops to negative values, which we attribute to a small, unavoidable initial polarization of the pump when focussed onto the sample. Second, $\rho_c$ does not drop to zero during the initial oscillations in the interval of 150-300\,ps. An intuitive explanation for this is the superposition of multiple oscillations with different frequency. This behavior can be understood in terms of the pseudospin model which will be discussed in depth in section \ref{chap:theory}.

\begin{figure}
	\includegraphics[width=1\columnwidth]{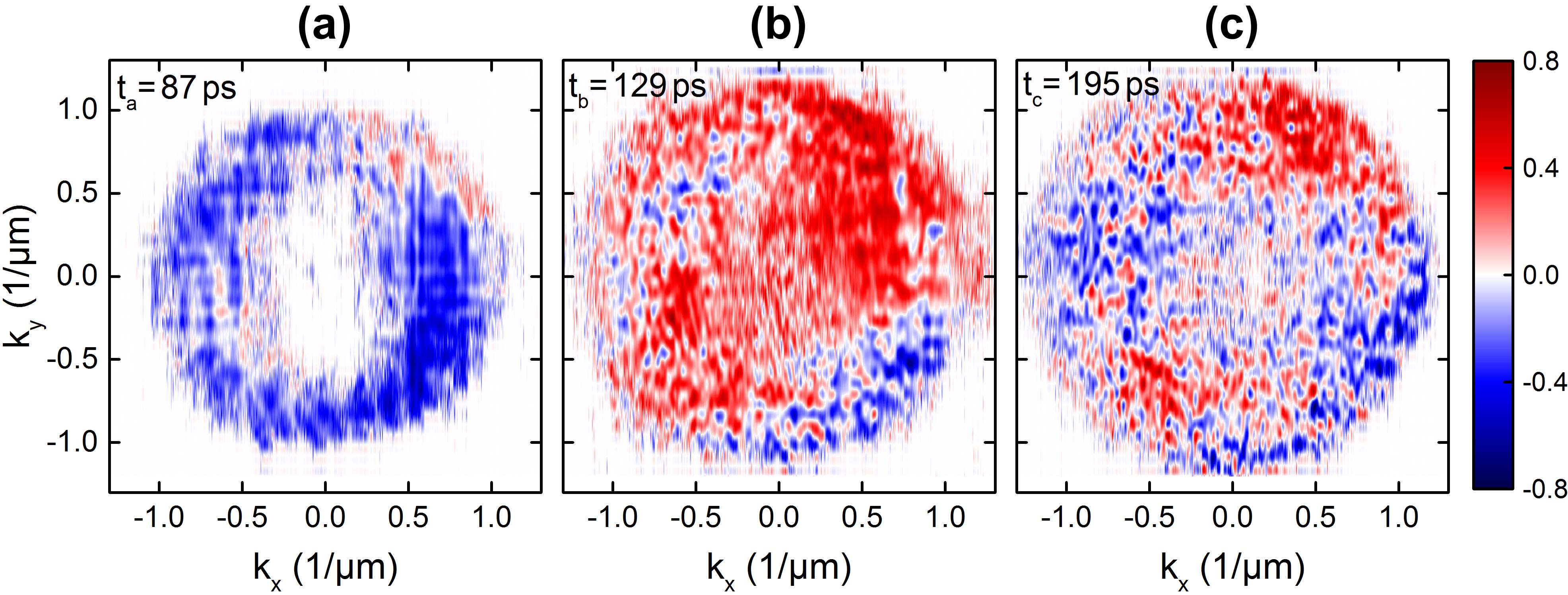}	
	\caption{$\rho_c$ in momentum space at selected points in time corresponding to the marked points in time of Fig. \ref{fig:dcp_time_integration_and_evolution}(b).}
	\label{fig:dcp_in_kspace_time_selection}
\end{figure}

Beyond the time evolution of characteristic points of the momentum space, some snapshots at certain points of interest in time merit further investigation, as shown in Fig. \ref{fig:dcp_time_integration_and_evolution}(b). $t_a$ coincides with the initial drop to a negative $\rho_c$. $t_b$ denotes the point in time with ma\-xi\-mal positive $\rho_c$. Finally $t_c$ is selected such that the overall $\rho_c$ of the total momentum space vanishes although in both areas $\rho_c$ is non-zero. The related momentum space snapshots are shown in Fig. \ref{fig:dcp_in_kspace_time_selection}. The first snapshot at $t_a=87$~ps shows a negative $\rho_c$ everywhere on the scattering ring. Since no precession is observable, we conclude that this is the time of excitation when resonant Rayleigh scattering without loss of polarization takes place. The conservation of polarization under resonant Rayleigh scattering has been observed before\cite{houdre2000,freixanet2000}. In the second snapshot in Fig. \ref{fig:dcp_in_kspace_time_selection} the total $\rho_c$ of the momentum space image shows the largest positive value and the scattering ring is dominated by a positive $\rho_c$, which also extends into the middle of the scattering ring due to the energy loss through ongoing scattering processes.
The main features of the typical OSHE pattern during the pseudospin precession still remain. The last snapshot in Fig. \ref{fig:dcp_in_kspace_time_selection} shows a rather finely segmented picture where points of similar $\rho_c$ are spread more widely in momentum space. Over the further course of the scattered beam this behavior stays the same until the intensity is too low to calculate reliable values of $\rho_c$.


\section{Theory}
\label{chap:theory}

Within the pseudospin formalism the polarization of microcavity polaritons is characterized by the three-component pseudospin vector $\mathbf{S} = (S_x,S_y,S_z)$, where $S_{x,y,z}$ describe the intensities of the linear (collinear with the original coordinate basis axes components and the diagonal/antidiagonal ones) and circular polarization components, respectively.
It directly maps to the conventional Stokes vector characterizing the polarization of the emitted light.

The evolution of the pseudospin in the scattered state $\mathbf{S}_{\mathbf{k}} \equiv \mathbf{S}_{\mathbf{k}}(t)$ characterized by the wave vector $\mathbf{k}$ can be described by the precession equation~\cite{leyder2007observation,kavokin2005,PhysRevB.77.165341}
\begin{equation}
\label{PrecEqnForPseudospin}
\frac{\partial \mathbf{S}_{\mathbf{k}}}{\partial t} = \mathbf{S}_{\mathbf{k}} \times \mathbf{\Omega}_{\mathbf{k}} + \mathbf{f}(t) - \frac{\mathbf{S}_{\mathbf{k}}}{\tau}
\end{equation}
accompanied by the rate equation for the population of the considered state
\begin{equation}
\label{RateEq}
\frac{\partial N_{\mathbf{k}}}{ \partial t} = f(t) - \frac{N_{\mathbf{k}}}{\tau}.
\end{equation}

The vector $\mathbf{\Omega}_{\mathbf{k}} = (\Omega_x,\Omega_y,\Omega_z)$ in the first term in the right-hand side of Eq.~\eqref{PrecEqnForPseudospin} describes the effective magnetic field causing oscillatory dynamics of the polariton pseudospin components.
In our model, the effective magnetic field represents a combination of three fields of different origin:
\begin{equation}
\label{EffMagnField}
\mathbf{\Omega _{k}} = \mathbf{\Omega} _{\mathbf{k}} ^{\text{LT}} + \mathbf{\Omega}_{\mathbf{k}}^{\text{an}} + \mathbf{\Omega}_{\mathbf{k}}^{\text{NL}}.
\end{equation}
The LT field $\mathbf{\Omega} _{\mathbf{k}} ^{\text{LT}}$ describes the effect of the LT splitting of the linear polarizations and is given by
\begin{equation}
\label{LTMagnFieldPart}
\mathbf{\Omega} _{\mathbf{k}} ^{\text{LT}} = \left[ \Delta_{\text{LT}} (k_x^2 - k_y^2), \, 2 \Delta_{\text{LT}} k_x k_y, \,0\right],
\end{equation}
where $\Delta_{\text{LT}}$ is the LT-splitting constant and
$k_{x,y}$ are the components of the wave vector characterizing the considered pseudospin state.
This splitting is mostly governed by the TE-TM splitting of the photonic microcavity mode~\cite{PhysRevB.59.5082} along with the long-range electron and hole exchange interaction~\cite{PhysRevB.47.15776}.
It is evident from Eq.~\eqref{LTMagnFieldPart} that the contribution of this field to the pseudospin dynamics is the stronger, the larger the absolute value of the wave vector $\mathbf{k}$ is.

The uniform effective field $\mathbf{\Omega} _{\mathbf{k}} ^{\text{an}} = [\delta _{\text{an}},0,0]$ characterizes the effect of the built-in anisotropy splitting arising from the optical and electronic anisotropy in the microcavity~\cite{Amo2009c,PhysRevB.82.085315}.
It is worth noting that the magnitude of this splitting is $\mathbf{k}$-independent.

While both $\mathbf{\Omega} _{\mathbf{k}} ^{\text{LT}}$  and $\mathbf{\Omega} _{\mathbf{k}} ^{\text{an}}$ affect the linear polarization components of $\mathbf{S} _{\mathbf{k}}$ in the cavity plane, the third effective magnetic field component, $\mathbf{\Omega} _{\mathbf{k}} ^{\text{NL}} = \left[0,0, \beta S_z\right]$, oriented in the growth direction plays the role of the effective Zeeman splitting of the circularly polarized polariton states.
It emerges in~Eq.~\eqref{PrecEqnForPseudospin} to characterize nonlinear effects during the polariton propagation.
Since the polaritons are excited resonantly, the dominant sources of nonlinearity are polariton-polariton interactions.
Accordingly, the effective field $\mathbf{\Omega} _{\mathbf{k}} ^{\text{NL}}$  describes the so-called self-induced Larmor precession of the polariton pseudospin~\cite{Flayac1367SS2630SS14SS8SS085018,PhysRevB.73.073303}.
The parameter $\beta$ is the effective polariton-polariton interaction constant that takes into account the difference of the exchange interaction strengths between two polariton states in the triplet configuration (parallel spins) and the singlet configuration (opposite spins).

The second term on the right-hand side of Eq.~\eqref{PrecEqnForPseudospin} describes the inflow of polaritons to the state $\mathbf{S_{k}}$.
Hereafter we shall account only for the elastic (Rayleigh) scattering of polaritons. 
We consider the originally studied con\-figu\-ra\-tion of the OSHE~\cite{kavokin2005} neglecting multiple scattering effects and assume the state $\mathbf{S_{k}}$ to be fed due to the scattering from the initial state $\mathbf{k}_0$ with $|\mathbf{k}_0|=|\mathbf{k}| = k$.
Within this model, the polariton flux is found as~\cite{kavokin2005,morina2013}
\begin{equation}
\label{FluxFromInitState}
\mathbf{f}(t) = 2 \frac{\mathbf{S}_{\mathbf{k}0}}{\tau _1} e^{-t/\tau},
\end{equation}
where $\tau_1$ is the scattering time describing the scattering of polaritons from the initial state $\mathbf{k}_{0}$ to the state $\mathbf{k}$.
$\mathbf{S}_{\mathbf{k}0}$~is~the pseudospin of the initial state.
The temporal evolution of the initial state can be found from the equation  $\partial \mathbf{S}_{\mathbf{k}0} / \partial t = \mathbf{S}_{\mathbf{k}0} \times \mathbf{\Omega} _{\mathbf{k}0}$, where the time dependence of the nonlinear component of the effective magnetic field, $\mathbf{\Omega} _{\mathbf{k}0} ^{\text{NL}}$, is taken into account in the explicit form.

The last terms in both~\eqref{PrecEqnForPseudospin} and~\eqref{RateEq} describe the radiative relaxation due to the finite polariton lifetime $\tau$ in a given state.

\begin{figure}
\includegraphics[width=1\columnwidth]{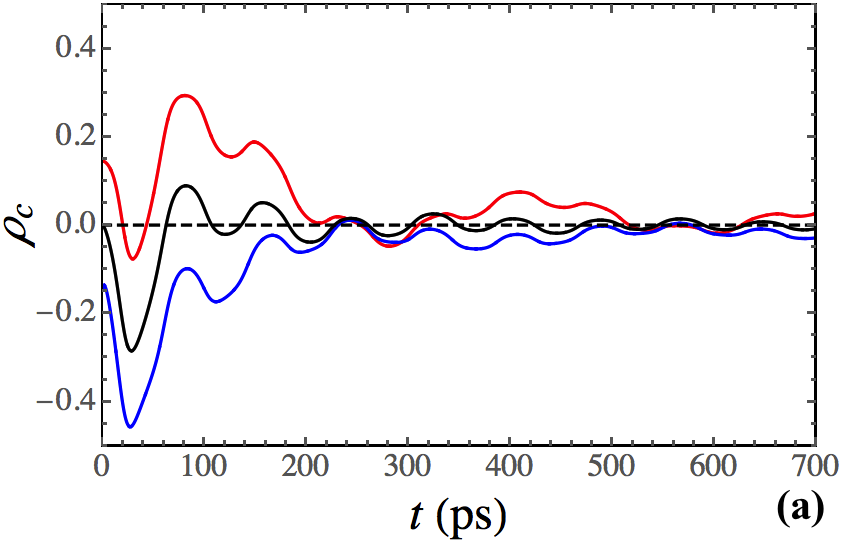}
\includegraphics[width=1\columnwidth]{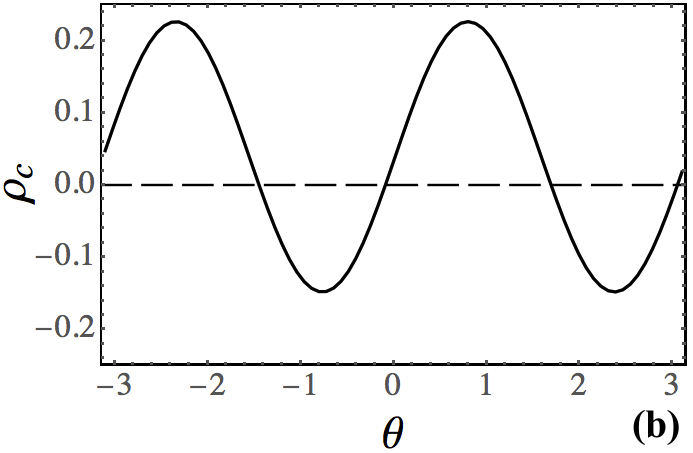}
\caption{
(a) Theoretical prediction of the time evolution of $\rho_c$ of the two polariton polarization states at angles of $+45^\circ$ (the red curve) and $-45^\circ$ (the blue curve) with respect to the direction of the pump.
The solid black curve corresponds to the total $\rho_c$ integrated over the elastic circle with $k=0.73\, \mu\text{m}^{-1}$.
(b) Integrated  $\rho_c$  against the scattering angle~$\theta$.
The parameters used for modelling are:
The LT splitting constant is $\hbar \Delta _{\text{LT}} = 18 \, \mu\text{eV} \times \mu \text{m}^2$,
the anisotropy splitting is $\hbar  \delta _{\text{an}} = 15 \, \mu\text{eV} $,
the effective interaction strength is $\hbar \beta =-0.45 \, \text{meV} \times \mu\text{m}^2$.
The lifetime of polaritons is
$\tau = 5$~ps and the scattering time is
$\tau _1 = 40$~ps.
}
	\label{fig:dcp_theor}
\end{figure}

Figure~\ref{fig:dcp_theor}(a) demonstrates the time evolution of $\rho_c$ of the two polariton polarization states corresponding to those in Fig.~\ref{fig:dcp_time_integration_and_evolution}(b).
The scattered state wave vector components are $k_x=k/\sqrt{2}$ and $k_y=\pm k/\sqrt{2}$ (``$+$'' and ``$-$'' correspond to the two observed states scattered at $+45 ^{\circ}$ and $-45 ^{\circ}$ with respect to the direction of the pump).
Both the dynamics of $\rho_c$ observed in Fig.~\ref{fig:dcp_time_integration_and_evolution}(b) and predicted in Fig.~\ref{fig:dcp_theor}(a) differ significantly from that calculated in Ref.~\onlinecite{kavokin2005} where the temporal dependencies of $\rho_c$ for the signal scattered at $+45^\circ$ and $-45^\circ$ are characterized by symmetric oscillations with a constant frequency which decay with some characteristic time.
This discrepancy arises due to the different geometry and initial conditions of our experiment with respect to Refs.~\onlinecite{kavokin2005,leyder2007observation}.
In particular, we need to take into account a slight ellipticity of the pump as follows:
$S_{\mathbf{k}0,x} (0)=-0.9$, $S_{\mathbf{k}0,y}  (0)=(1-S_{\mathbf{k}0,x}^2)^{-1/2}$ and $S_{\mathbf{k}0,z}  (0)=0$.
The ellipticity is caused by the optical setup we have used.
In addition, in our case $\rho_c$ is not zero at the beginning of the observed time frame.
By choosing the initial conditions $\mathbf{S}_{\mathbf{k}} (0)$ that are fitting parameters of the model one can achieve a better agreement between the modelling results and the experimental data.

According to Fig.~\ref{fig:dcp_theor}(a), $\rho_c$ undergoes oscillatory dynamics, herewith one can distinguish two kinds of oscillations.
The slower oscillations with a period of about 300 picoseconds originate from the pseudospin precession around the effective magnetic field with the oscillation frequency given by $\Omega = |\mathbf{\Omega}_{\mathbf{k}}|$.
The fast ones with a period of about 70-80 picoseconds are caused by the mismatch of the oscillation frequencies of the scattered state and the pump state due to the nonlinear effect of polariton-polariton interactions.

Figure~\ref{fig:dcp_theor}(b) shows the time-averaged distribution of $\rho_c$ around the elastic circle corresponding to the incident pump wave vector $\mathbf{k}_0$. 
The dependence of $\rho_c$ on the scattering angle $\theta$ demonstrates a periodic character.
However, in contrast to the predictions of~Ref.~\onlinecite{kavokin2005}, the positions of the maxima are shifted with respect to the zero-angle direction, $\theta = 0$ which manifests the rotation of the distribution of $\rho_c$ in the QW plane.
Another interesting difference is that the distribution is biased towards positive values of $\rho_c$.
The origin of both effects can be traced back to the impact of the $z$-component of the effective magnetic field induced by the spin-dependent polariton-polariton interactions.
A similar contribution of the $z$-component to the distribution of $\rho_c$ has been predicted theoretically in Ref.~\onlinecite{morina2013} for an external magnetic field applied normal to the cavity plane.

\section{Conclusion}
We have experimentally demonstrated the time behavior of the OSHE.
The observed process can be described within the framework of the pseudospin model developed assuming a single Rayleigh scattering act per propagating polariton and taking into account the effect of the self-induced Zeeman splitting of polariton states due to spin-dependent polariton-polariton interactions.
The polariton pseudospin undergoes a characteristic precession around the effective magnetic fields of different origin that results in the oscillatory behavior of $\rho_c$ of the emitted light.

\section{Acknowledgements}
We gratefully acknowledge the financial support by the Deutsche Forschungsgemeinschaft in the frame of the ICRC TRR 160 within project B7.
The W\"urzburg group acknowledges the support by the Deutsche Forschungsgemeinschaft within the project SCHN1376-3.1.
E.S. acknowledges support from the Russian Foundation for Basic Research grant No.~16-32-60104.
A.K. and E.S. acknowledge support from the EPSRC Hybrid Polaritonics Programme grant.
A.K. acknowledges the partial support from the HORIZON 2020 RISE project CoExAn (Grant No.~644076).


%

\end{document}